\documentclass[conference]{IEEEtran}
\IEEEoverridecommandlockouts
\usepackage{amsmath,amssymb,amsfonts}
\usepackage{algorithmic}
\usepackage{graphicx}
\usepackage{textcomp}
\usepackage{float}
\usepackage{comment}
\usepackage{float}
\usepackage{multirow}
\usepackage{caption}
\usepackage{subcaption}
\usepackage{academicons}
\usepackage{placeins}
\usepackage{hyperref}
\usepackage{xcolor} 
\usepackage{xcolor} 
\definecolor{orcidlogocol}{HTML}{A6CE39} 

\usepackage[style=numeric, backend=biber]{biblatex}

\def\BibTeX{{\rm B\kern-.05em{\sc i\kern-.025em b}\kern-.08em
    T\kern-.1667em\lower.7ex\hbox{E}\kern-.125emX}}
    
\begin{document}

\title{Beyond SSO: Mobile Money Authentication for Inclusive e-Government in Sub-Saharan Africa}

\author{
\IEEEauthorblockN{
Oluwole Adewusi,
Wallace S. Msagusa,
Jean Pierre Imanirumva, 
Okemawo Obadofin,
Jema D. Ndibwile,
\IEEEauthorblockA{College of Engineering, Carnegie Mellon University Africa, Kigali, Rwanda \\
Emails: \{oadewusi, wmsagusa, jimaniru, oobadofi, jndibwil\}@andrew.cmu.edu. \\
*Corresponding author.}
}
}

\maketitle

\begin{abstract}

The rapid adoption of Mobile Money Services (MMS) in Sub-Saharan Africa (SSA) offers a viable path to improve e-Government service accessibility in the face of persistent low internet penetration. However, existing Mobile Money Authentication (MMA) methods face critical limitations, including susceptibility to SIM swapping, weak session protection, and poor scalability during peak demand.

This study introduces a hybrid MMA framework that combines Unstructured Supplementary Service Data (USSD)-based multi-factor authentication with secure session management via cryptographically bound JSON Web Tokens (JWT). Unlike traditional MMA systems that rely solely on SIM-PIN verification or smartphone-dependent biometrics, our design implements a three-factor authentication model; SIM verification, PIN entry, and session token binding,  tailored for resource-constrained environments.

Simulations and comparative analysis against OAuth-based Single Sign-On (SSO) methods reveal a 45\% faster authentication time (8 seconds vs. 12–15 seconds), 15\% higher success under poor network conditions (95\% vs. 80\%), and increased resistance to phishing and brute-force attacks. Penetration testing and threat modeling further demonstrate a substantial reduction in vulnerability exposure compared to conventional approaches.

The primary contributions of this work are: (1) a hybrid authentication protocol that ensures offline accessibility and secure session continuity; (2) a tailored security framework addressing threats like SIM swapping and social engineering in SSA; and (3) demonstrated scalability for thousands of users with reduced infrastructure overhead. The proposed approach advances secure digital inclusion in SSA and other regions with similar constraints.

\end{abstract}


\section{Introduction}

In regions characterized by sparse computing infrastructure and intermittent internet connectivity, governments and organizations across Sub-Saharan Africa (SSA) face significant challenges in digitizing essential services such as e-Government platforms and the distribution of relief funds. This challenge is especially pronounced in rural and under-served communities where reliable internet access is limited, reducing the effectiveness of traditional web-based authentication solutions. Conversely, mobile phone penetration, particularly through basic devices capable of Unstructured Supplementary Service Data (USSD), remains widespread due to their simplicity, reliability, and low technical requirements. Given this technological landscape, stakeholders must develop secure, accessible, and user-friendly authentication mechanisms tailored to low-connectivity environments.

The widespread adoption of Mobile Money Services (MMS) in SSA represents an essential but underutilized infrastructure capable of addressing these authentication challenges. MMS have become integral to daily financial activities in countries such as Kenya, Tanzania, Rwanda, and Uganda, significantly enhancing financial inclusion through secure, widely adopted transaction platforms \cite{osabutey2024}. However, while MMS demonstrates substantial promise, existing mobile money-based authentication methods exhibit vulnerabilities, including susceptibility to SIM swapping and limited protection against phishing attacks. Additionally, conventional web-based Single Sign-On (SSO) authentication methods often fail to accommodate the infrastructural limitations and digital literacy challenges prevalent in the region. For instance, prior studies in Kenya revealed that e-Government users frequently rely on third-party internet café operators to create and manage their online accounts due to the stringent credential management requirements, compromising user privacy and security \cite{10179410}.

\begin{figure}[ht!]
    \centering
    \includegraphics[width=1.0\linewidth]{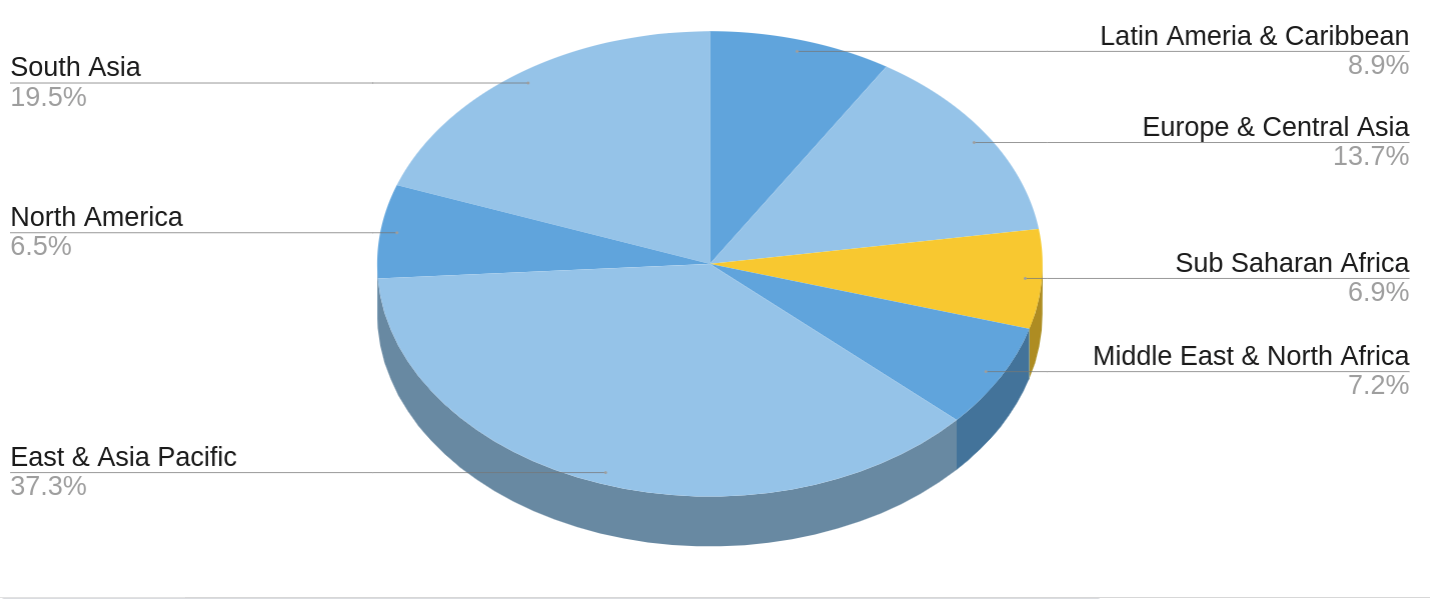}
    \caption{Mobile Internet connectivity by region, Global System for Mobile Communications Association (GSMA).}
    \label{fig:digital-divide}
\end{figure}

\vspace{5pt}
Mobile phone ownership in SSA reached 43\% by 2024, far surpassing internet access  \cite{gsma2024}, creating an opportunity to leverage mobile channels for inclusive authentication. Out of the 4.61 billion people with access to mobile internet worldwide, only 320 million (approx. 7\%) are from the SSA, as highlighted in Figure~\ref{fig:digital-divide}. This considerable disparity highlights the potential of mobile technology as a viable alternative for enabling secure digital services in low-connectivity areas. While web-based SSO often compromises security—especially where users rely on internet cafes to manage credentials \cite{10179410}—mobile money offers a trusted alternative. Widely adopted in East Africa, these platforms have embedded themselves into daily financial life and present an opportunity to build secure, infrastructure-aware authentication methods

This study addresses these issues by proposing a novel, practical, and technology-agnostic Mobile Money Authentication (MMA) framework that integrates USSD-based multi-factor authentication (SIM verification, PIN authentication, and secure session token binding) with robust session management using cryptographically secure JSON Web Tokens (JWT). Unlike existing solutions reliant solely on basic SIM-PIN verification or smartphone-dependent biometric authentication, our approach provides a hybrid authentication mechanism tailored explicitly for SSA's resource-constrained contexts.

The main contributions of this paper are:
(1) Identification and thorough analysis of critical authentication challenges specific to SSA, emphasizing the necessity for tailored solutions to address infrastructural and technological constraints.
(2) Development of a hybrid MMA approach integrating USSD technology and advanced session management to provide enhanced security, improved accessibility, and better user experience.
(3) Comprehensive evaluation and comparative analysis demonstrating MMA’s superior performance compared to conventional SSO methods across dimensions of security, accessibility, user experience, cost-efficiency, and scalability, particularly in environments with limited connectivity.

The remainder of the paper is structured as follows: Section II reviews existing work on mobile money systems and SSO methods. Section III presents the background and threat model tailored specifically to SSA. Section IV details the design and implementation of the proposed MMA framework. Section V provides experimental results and comparative analysis. Section VI discusses the limitations and future research directions, and Section VII concludes with a summary of findings and their broader implications.

\section{Literature Review}  
This section explores existing research on authentication methods, with a focus on the strengths and limitations of SSO and MMS. Secure and accessible authentication solutions have become increasingly critical, particularly in regions like SSA where internet connectivity and computing resources are limited. By examining the security and user experience aspects of SSO and MMS, this review highlights the challenges associated with authentication methods and the potential of MMA to address these gaps. Furthermore, this section includes a detailed exploration of MMS as a specific and transformative use case, focusing on their architecture, security considerations, and their pivotal role in financial inclusion.

\vspace{2pt}
\subsection*{\textbf{Mobile Money Services(MMS)}}
\addcontentsline{toc}{subsection}{Mobile Money Services(MMS)}
\vspace{-3pt}

\begin{figure}
    \centering
    \includegraphics[width=1.0\linewidth]{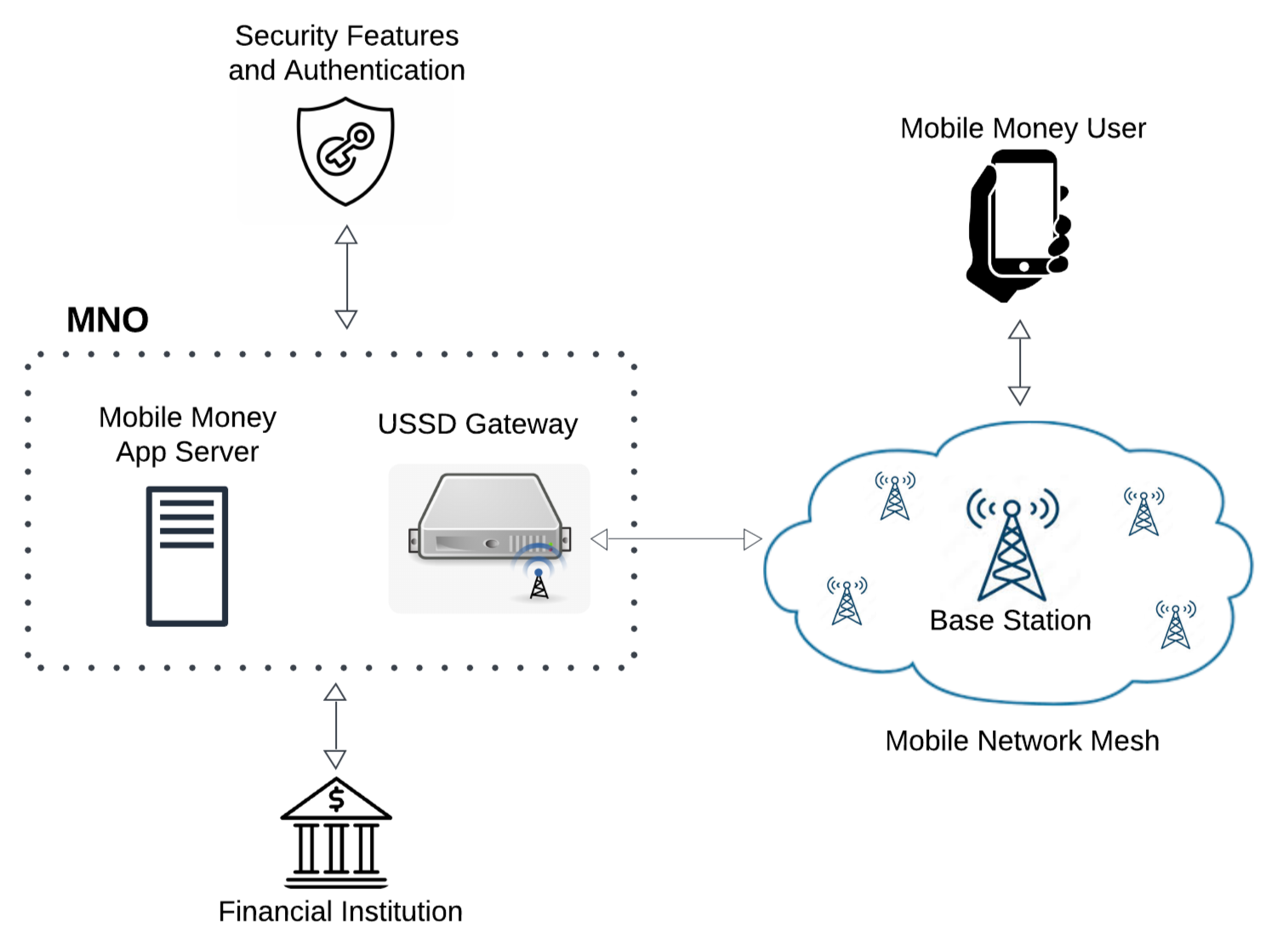}
    \caption{Architecture of Mobile Money.}
    \label{mobile_money}
\end{figure}

Mobile money has transformed financial inclusion in developing regions by allowing underbanked populations to access financial services through mobile devices. The standardized architecture—comprising of user devices, Subscriber Identity Module (SIM) cards, Mobile Network Operators (MNOs), USSD gateways, and financial institutions—creates a secure framework that can be leveraged beyond financial transactions for authentication purposes.  Figure ~\ref{mobile_money} provides a visual representation of this architecture, adapted from the standard design discussed in the comprehensive review of MMA \cite{aliguma2020}.

Historically, MMS gained traction in SSA due to the limited banking infrastructure. Systems such as M-PESA, operated by Safaricom in Kenya and Vodacom in Tanzania, and MoMo, provided by MTN in countries like Rwanda and Uganda, revolutionized financial transactions by enabling users to send, receive, and store money digitally. Access to these services often occurs via USSD protocols, which allow users to interact with the platform through basic mobile phones without internet access.

\vspace{2pt}
\subsection*{\textbf{Web Security of SSO vs. MMA}}
\vspace{-3pt}
SSO systems have gained widespread popularity since their inception, offering users the convenience of logging in once and accessing multiple web services seamlessly. However, this convenience comes with significant challenges, particularly in regions such as SSA, where stable internet connectivity and robust cybersecurity infrastructure are often lacking. In contrast, SSA stands out for its limited access to traditional banking services but widespread adoption of MMS, which presents a transformative opportunity in the digital landscape.

SSO systems, as highlighted by Bazaz and Khalique \cite{bazaz2016}, simplify credential management but depend heavily on stable internet connections and centralized servers. These dependencies create vulnerabilities in regions with unreliable connectivity and heightened cybersecurity risks. In contrast, MMS, promoted by Hossain et al. \cite{hossain2018}, operate securely through offline channels such as USSD, allowing financial transactions without the need for internet access. This makes MMA a compelling alternative for areas where reliability and accessibility are critical.

MMA employs robust security mechanisms, including Personal Identification Number (PIN) codes and SIM card verification, which significantly enhance defenses against phishing and unauthorized access. In contrast, OAuth-based SSO systems, while praised for their simplicity and user-friendly design, are susceptible to vulnerabilities such as cross-site request forgery (CSRF) and cross-site scripting (XSS) attacks. These risks are particularly concerning in low-income regions that lack advanced cybersecurity measures \cite{hossain2018}. Using MMA, these challenges can be mitigated, offering a secure and inclusive alternative tailored to the unique needs of resource-constrained environments.

\vspace{2pt}
\subsection*{\textbf{Challenges of SSO Systems and Mobile Money Solutions}}  
\vspace{-3pt}
 SSO technologies, as noted by Bazaz and Khalique \cite{bazaz2016}, face significant limitations in developing countries due to their dependence on stable internet connections and centralized servers. These dependencies expose SSO systems to risks associated with unreliable infrastructure and increased susceptibility to cyber threats. In contrast, MMA uses technologies such as USSD and SMS, avoiding these vulnerabilities by utilizing the ubiquity of mobile networks. However, as Kibuuka \cite{kibuuka2020} highlights, MMS are not without challenges, particularly security concerns such as SIM swap fraud and Man-in-the-Middle (MITM) attacks. Despite these risks, mobile money remains a critical enabler of financial inclusion in regions with limited internet connectivity, underscoring its potential for secure authentication solutions.

Innovative strategies are essential to optimize the implementation of MMA for web platforms. Biometric methods, as explored by Edoh et al. \cite{edoh2023}, offer enhanced security but are often hampered by high implementation costs and technological barriers in low-resource settings. Blockchain-based solutions, while promising for decentralized security, face practical challenges in scalability and usability in resource-constrained environments \cite{mamdouh2021}. Adaptive authentication systems, as proposed by Abu et al. \cite{abu2014}, provide flexibility by tailoring security requirements to user behavior. However, their inherent complexity and resource demands may make them impractical in contexts where simplicity, affordability, and accessibility, the inherent qualities of MMA, are paramount. These considerations emphasize the need for balanced approaches that leverage mobile money's strengths while addressing its limitations to create secure and inclusive authentication solutions.
\vspace{2pt}
\subsection*{\textbf{Privacy-Preserving MMA}}  
\vspace{-3pt}
Federated identity management presents significant challenges in establishing trust relationships and implementing robust security measures between organizations, as highlighted by Jensen et al. \cite{jensen2011}. In contrast, MMA offers a localized and scalable solution that seamlessly integrates with existing mobile infrastructures. In particular, about 58\% of MMS are hosted by MNOs, leveraging accessible technologies such as USSD and SMS to ensure widespread adoption and ease of use \cite{GSMA2015SOTIR}. With growing concerns about data privacy, Ioannou et al. \cite{ioannou2020} stress the importance of privacy-preserving authentication systems that protect user identities throughout the authentication process. MMA addresses these concerns by directly employing encrypted communication channels that maintain data integrity and foster user trust \cite{adam2015}. In the context of e-Government platforms, Tennakoon \cite{tennakoon2020} highlights the unique challenges facing developing countries, where balancing accessibility and security is crucial. MMA emerges as a viable and effective solution, offering a secure, scalable, and user-friendly method tailored to the specific needs and constraints of these regions.

\vspace{5pt}

Research on mobile money has predominantly focused on enhancing financial inclusion and service accessibility. USSD technology facilitates financial activities such as bill payments, fund transfers, and balance inquiries \cite{osabutey2024}. Governments have leveraged USSD for public money transfers during emergencies, underscoring its reliability and accessibility \cite{marin2022digital}. Innovations like Tunisia’s E-Wallet, linked to the national digital identity system E-Houwiya, enable secure payments and withdrawals over USSD \cite{benneji2023tunisia}. However, these applications primarily address financial services rather than web authentication. Our research introduces an architecture that leverages mobile money security features, such as SIM verification and PIN-based authentication, to enhance web platform security. This approach is particularly suitable for regions with high mobile phone penetration and limited internet connectivity, offering a cost-effective and inclusive solution based on familiar technology.

\section{Background and Security Analysis}
This section provides foundational context and establishes a structured security framework for our proposed MMA approach. We begin by highlighting the critical role e-Government platforms play in fostering digital transformation across SSA, particularly in addressing governance challenges and improving service delivery. Despite the promise of these platforms, their effective deployment has been limited by significant barriers, including intermittent connectivity, device limitations, and varying levels of digital literacy among users.

The unique sociotechnical landscape in SSA introduces specific vulnerabilities, making SSO authentication methods inadequate. SSO authentication paradigms often fail to accommodate intermittent network access, reliance on shared or low-capability devices, and susceptibility to social engineering due to uneven cybersecurity awareness. These factors necessitate a tailored security approach that explicitly considers regional technological constraints and user behavior patterns.

To address these contextual vulnerabilities, we developed a threat model aligned with SSA’s technological environment. Our model identifies potential threats, including SIM swapping, phishing attacks, data interception, session hijacking, denial of service, and unauthorized access escalation. Using the STRIDE framework, which encompasses threats related to spoofing, tampering, repudiation, information disclosure, denial of service, and elevation of privilege, we categorize and articulate clear security requirements.

Our threat model makes realistic assumptions regarding adversarial capabilities, considering that attackers might possess the means to intercept network communications, exploit publicly available user data, or conduct social engineering attacks prevalent in the region. Conversely, we assume attackers cannot breach a robust encryption standard such as AES-256, simultaneously compromise multiple authentication factors (SIM card and PIN), or infiltrate both telecommunications and authentication infrastructures at a large scale.

Building upon these threat considerations, our MMA-based authentication design emphasizes critical security properties: authentication through multi-factor mechanisms (SIM + PIN), confidentiality via robust encryption, integrity ensured by session token validation, non-repudiation through comprehensive audit logging, and sustained availability even under adverse network conditions. Our subsequent technical implementation (Section IV) integrates these considerations explicitly, ensuring MMA addresses both the functional and security imperatives identified here.

\vspace{2pt}
\subsection*{\textbf{Why Focus on e-Government Platforms}}  
\vspace{-3pt}
E-Government platforms are central to global digital transformation efforts, offering a streamlined and efficient medium to deliver essential government services to citizens. These platforms span a wide array of services, including taxation, social security, licensing, healthcare, and voter registration, among others. In SSA, where rapid digitization is crucial to improving governance and citizen engagement, e-Government platforms have become indispensable tools to bridge the gaps in service delivery. Several key factors underscore the importance of focusing on these platforms.  

E-Government systems serve as a cornerstone of digital transformation, enabling governments to automate vital services, reduce bureaucracy, and foster transparency. By providing citizens with direct access to services, these platforms eliminate the need for physical visits to government offices, addressing challenges such as long waiting times, inefficiency, and corruption.  

In SSA, where internet penetration remains limited compared to the widespread adoption of mobile phones, e-Government platforms hold significant potential to bridge the digital divide. Integrating MMA into these platforms can extend their reach to under-served populations, particularly impoverished and rural communities. Such integration ensures digital inclusion by enabling access to essential services for citizens who lack traditional banking infrastructure or reliable internet connectivity—a transformative and attainable goal for the region.

\vspace{2pt}

\subsection*{\textbf{Threat Model and Security Considerations}}
\vspace{-3pt}
To ensure the robustness of our MMA approach, we developed a threat model addressing the unique security challenges in SSA's technological landscape. This model identifies potential attack vectors and establishes security requirements for e-Government authentication in resource-constrained environments.

\subsection{Adversarial Capabilities and Assumptions}
\vspace{-3pt}
We considered adversaries with various capabilities within the SSA context. These include potential network traffic interception, access to publicly available user information (including phone numbers), temporary physical access to shared devices (a common scenario in SSA), and social engineering skills leveraging regional vulnerabilities.
However, our model assumes certain limitations on attacker capabilities. We presumed attackers lack the cryptographic capabilities to break properly implemented AES-256 encryption and cannot simultaneously compromise both authentication factors (the SIM card and PIN knowledge). Additionally, we assumed they cannot compromise core telecommunications infrastructure at scale nor simultaneously breach both the authentication server and MNO systems.

\subsection{Threat Categories (STRIDE Framework)}

Our threat model uses the STRIDE framework to categorize potential vulnerabilities:

\noindent\textbf{Spoofing} refers to threats where attackers impersonate legitimate users or system components. In the SSA context, this includes SIM swapping attacks, phishing attempts to obtain PIN codes, and impersonation of e-Government portals or USSD gateways.

\noindent\textbf{Tampering} involves unauthorized modification of data, such as manipulation of authentication data in transit, unauthorized modification of session information, and tampering with USSD requests or responses.

\noindent\textbf{Repudiation} concerns threats related to users denying their actions. These may include unauthorized transactions with plausible deniability, disputes over authentication attempts, and lack of non-repudiation for critical government service access.

\noindent\textbf{Information Disclosure} encompasses threats exposing sensitive data, including personal identification information, authentication credentials, and unauthorized access to government service data.

\noindent\textbf{Denial of Service} includes threats preventing legitimate system use, such as USSD gateway flooding, telecommunications network congestion attacks, and authentication server overloading.

\noindent\textbf{Elevation of Privilege} relates to threats allowing unauthorized access levels, including unauthorized access to administrative functions, authentication mechanism bypasses, and exploitation of session management vulnerabilities.

\subsection{Context-Specific Threats in SSA}

Beyond standard threat categories, we identified region-specific concerns. Infrastructure reliability threats include intermittent network connectivity that may compromise session integrity, power outages affecting authentication completion, and limited cellular coverage in rural areas impacting service availability.
Digital literacy challenges present additional security concerns, such as PIN sharing among family members due to limited technology access, susceptibility to social engineering due to varying levels of cybersecurity awareness, and difficulty distinguishing legitimate from fraudulent authentication requests.
Device-level constraints further complicate security in SSA, with shared device usage in community settings, feature phones with limited security capabilities, and offline device usage patterns all requiring specific security considerations.

\subsection{Security Properties and Mitigations}

Our MMA system is designed to achieve several key security properties: authentication (verifying user identity through SIM card and PIN), confidentiality (protecting sensitive data), integrity (ensuring unaltered data transmission), non-repudiation (providing evidence of authentication actions), and availability (ensuring service accessibility despite connectivity challenges).
To address spoofing threats, we implemented multi-factor authentication combining SIM verification and PIN, account lockout after multiple failed attempts, and out-of-band authentication using separate communication channels.
For tampering and information disclosure threats, our system employs end-to-end encryption for sensitive data, secure session management using JWTs with appropriate expiration, and minimal data collection and transmission principles.
Repudiation threats are mitigated through comprehensive logging of authentication events, transaction receipts via SMS, and secure audit trails. Denial of service protection includes rate limiting on authentication attempts, distributed system architecture, and graceful degradation under limited connectivity.
Finally, to prevent elevation of privilege, we implemented the principle of least privilege in system design, regular security audits, and secure session management practices appropriate for the SSA context.
This threat model provides the foundation for our security approach and informs the specific implementation choices and configurations described in the remainder of this paper.

Given the alignment between the challenges of SSA and the capabilities of MMA, we propose its integration into e-Government platforms as a scalable, secure, and inclusive solution to enhance trust, service delivery, and accessibility
\section{System Design and Methodology} 
To ensure reproducibility and transparency, this section provides a clear description of the experimental configuration, simulation parameters, and testing methodology used for our comparative analysis between MMA and  Single Sign-On (SSO) authentication methods.
\vspace{2pt} 
\subsection*{\textbf{Hardware and Software Environment}}
The simulation environment consisted of multiple dedicated servers, each configured specifically for its functional role. The authentication server operated on Ubuntu 22.04 LTS (4 vCPUs and 8GB RAM). A web server with an identical operating system but enhanced resources (4 vCPUs, 16GB RAM) hosted the user-facing components, while a similarly configured database server supported backend data storage and management. Additionally, a simulated Mobile Network Operator (MNO) server running Ubuntu 22.04 (2 vCPUs, 4GB RAM) was deployed to handle USSD interactions.

We built the system using Node.js (v18.15.0), Express (4.18.2), React (v18.2.0), JDK 22 and JWT authentication, with MongoDB and PostgreSQL for data storage. A custom Spring Boot application simulated USSD authentication, while Locust handled load testing. Testing covered various devices (feature phones to smartphones) and browsers (Chrome, Firefox, Safari). For realistic SSA conditions, we created three connectivity profiles based on telecommunications reports \cite{gsma2024}\cite{worldbank_digital_2021}: Good (4G/LTE: 15Mbps download, 50ms latency), Moderate (3G: 2Mbps download, 150ms latency), and Poor (2G/EDGE: 250Kbps download, 300ms latency).

In contrast, the USSD-based MMA interactions relied on Global System for Mobile communications (GSM) signaling channels rather than internet-based communication. For USSD, network quality primarily affects latency, response times, and session reliability rather than bandwidth. Therefore, we separately simulate USSD conditions characterized by varying degrees of GSM network stability: stable (urban), intermittent (rural), and poor (remote) to evaluate how these constraints of GSM signaling influence the performance and reliability of MMA authentication. It is important to note that while USSD does not rely on internet protocols, our simulation includes degraded network conditions to reflect the broader system's hybrid interaction with web-based APIs, particularly in mobile money web portal scenarios. Network impairments, such as packet loss, jitters, and random disconnections, were introduced using Linux Traffic Control (tc) tools to make the simulation more realistic.

\vspace{-3pt}
\noindent
\subsection{Implementation of Authentication Mechanisms}
\vspace{-3pt}
For MMA, we implemented a custom USSD gateway simulation following standard protocols. This included realistic USSD menu flows matching prevalent regional MMS, secure PIN verification algorithms, and robust session management through securely generated tokens. The web platform interacted with these simulations through RESTful APIs, managing JWT issuance, validation, session handling, and appropriate CORS configurations.

\noindent
For  SSO methods serving as a baseline, we employed an OAuth 2.0-compliant implementation with OpenID Connect capabilities. This included email/password authentication with optional email-based two-factor authentication and standard OAuth 2.0 authorization flows. In addition, social login integrations using Google's OAuth and Facebook Login were implemented for comparative testing.

\vspace{-3pt}
\noindent
\subsection{Testing Methodology}
\vspace{-3pt}
A structured evaluation methodology was used to ensure reliable, comparable outcomes. Authentication flows were tested through scripted user journeys, systematically measuring the time required to authenticate and tracking success rates across different network conditions. Each authentication method underwent 500 attempts per network scenario to guarantee statistically significant results.

Load testing involved simulating concurrent user sessions (50, 100, and  500 users) with a one-minute ramp-up to peak load, sustained for fifteen minutes. Performance metrics recorded included response times, error rates, and resource usage metrics (CPU and memory consumption).
Security assessments were thorough, employing automated OWASP ZAP vulnerability scanning and manual penetration tests aligned with OWASP guidelines. Specific attention was given to brute-force attack resilience through rate-limited simulations and session hijacking attempts. Network resilience was evaluated by performing authentication during simulated network interruptions and measuring recovery behaviors and session persistence under fluctuating connectivity.

\vspace{-3pt}
\noindent
\subsection{Metrics and Statistical Validation}
\begin{figure}
\centering
\includegraphics[width=\linewidth]{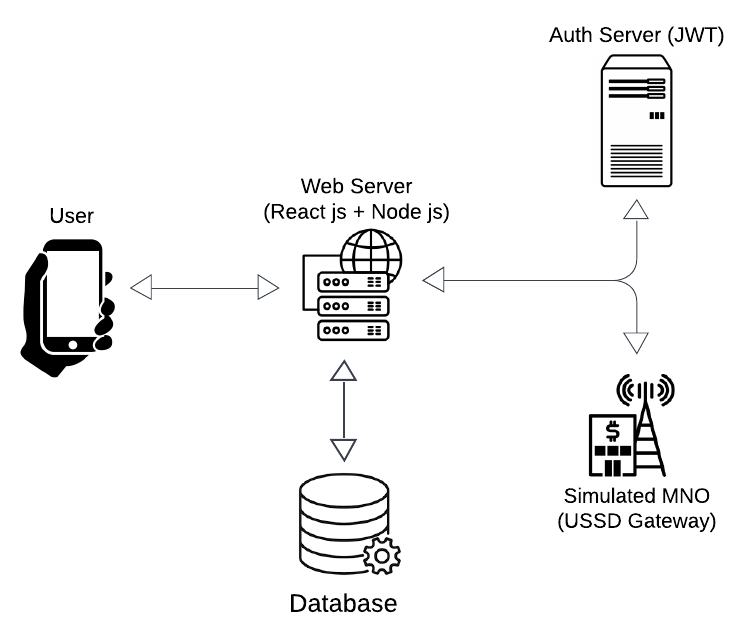}
\caption{Web Platform Architecture for Mobile Money Authentication (MMA).}
\label{fig:architecture}
\end{figure}

\vspace{-3pt}
Primary metrics included authentication times (seconds), success rates (\%), security incident rates (\%), and user effort (interaction counts). Statistical rigor was ensured by determining the minimum required sample size through power analysis ($\alpha=0.05$, $\beta=0.2$). Each test was repeated at least 30 times to achieve reliable statistical significance, with confidence intervals (95\%) calculated for all reported metrics. Differences between authentication methods were evaluated using two-tailed t-tests, and Cohen's $d$ measured the effect sizes.
\vspace{2pt} 
This methodology ensures thoroughness, transparency and reproducibility, facilitating direct replication and validation of our results by other researchers or practitioners.
To replicate the MMS in our simulation, we recreated its critical components and configured them to reflect real-world conditions. Using the standard architecture for mobile money, as shown in Figure~\ref{mobile_money}, we developed a web-based platform that accurately mirrors the essential processes and interactions involved in the authentication of mobile money. Designed to simulate an e-Government service, the platform integrates multiple components, including user interfaces, an authentication server, and a simulated MNO, enabling secure and efficient user authentication. As illustrated in Figure ~\ref{fig:architecture}, the architecture ensures seamless interactions between users, the e-Government platform, and the simulated mobile money system, providing a robust framework for testing and analysis.

\vspace{2pt}
\subsection*{\textbf{Core System Components}}
\begin{itemize}

\item Frontend Interface (React.js): Serves as the primary user interaction point
\item Backend Server (Spring Boot): Coordinates user requests and manages sessions
\item Authentication Server (Node.js): Issues and validates security tokens
\item Database (MongoDB \& PostgreSQL): Stores user data and session information
\item Simulated MNO (Node.js, \& React Native): Replicates USSD interactions for authentication. 
\end{itemize}
Authentication is managed by a dedicated server that issues JWTs upon successful credential verification. JWTs are used for secure session management, maintaining tamper-proof communication across all components. The back-end integrates with a simulated MNO through a USSD gateway to replicate real-world MMA. When a user initiates a request, the back-end triggers a USSD notification via the MNO, prompting the user to enter their PIN. The PIN is verified by the MNO, and the result is securely relayed back to the back-end through the authentication server. To ensure robust security, all communications use AES-256 encryption, while the modular architecture supports scalability to accommodate a growing user base and additional services.

\subsection*{\textbf{Authentication Flow}}

\begin{figure}[htbp]
\centering
\includegraphics[width=1.0\linewidth]{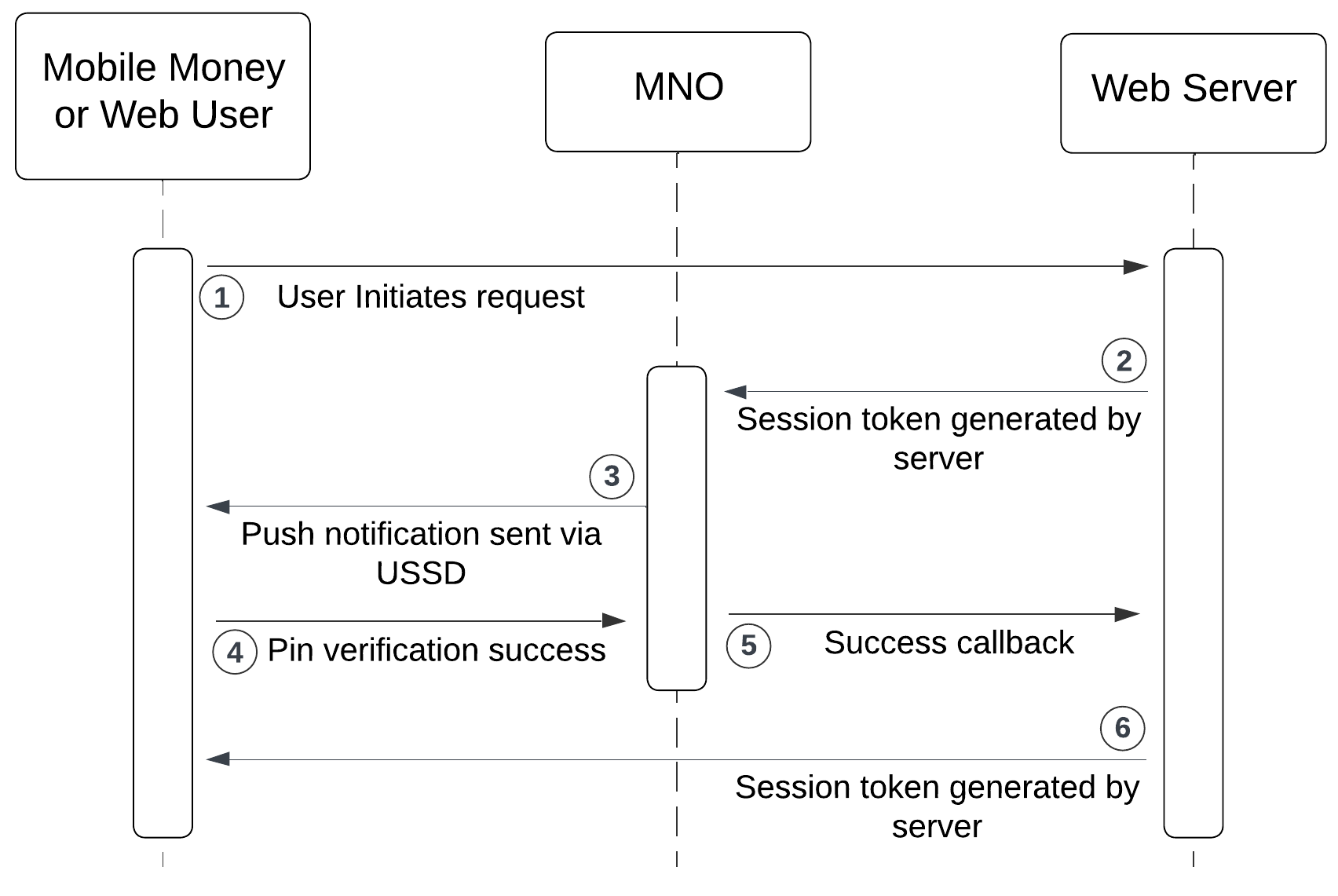}
\caption{Authentication Sequence.}
\label{fig:authentication_flow}
\end{figure}

\vspace{-3pt}
The authentication process, as shown in Figure ~\ref{fig:authentication_flow}, begins with a user initiating a registration or log-in request through the platform's front-end interface. This request is forwarded to the backend server, which orchestrates the subsequent steps to authenticate the user.

\begin{figure}[htbp]
    \centering
    \includegraphics[width=1.00\linewidth]{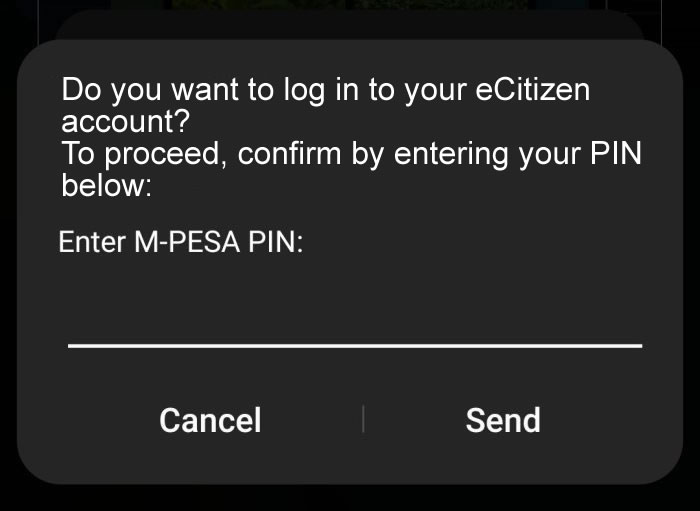}
    \caption{Sample Kenya M-Pesa Prompt for Our Architecture.}
    \label{fig:prompt}
\end{figure}

Upon receiving the request, the backend server generates a unique session token and communicates with the simulated MNO. The MNO sends a USSD push notification (Figure~\ref{fig:prompt} to the user's mobile number, offering options to authenticate with their PIN or deny the request.

 If the user enters their PIN, it's verified by the MNO and authentication server. Upon successful verification, a JWT is generated and transmitted to the backend server, which grants access to the requested service. This approach enables secure authentication even without internet access.

\vspace{2pt}
\subsection*{\textbf{Mock Interface MMA}}
\begin{figure*}[htbp]
    \centering
    \includegraphics[scale=0.35]{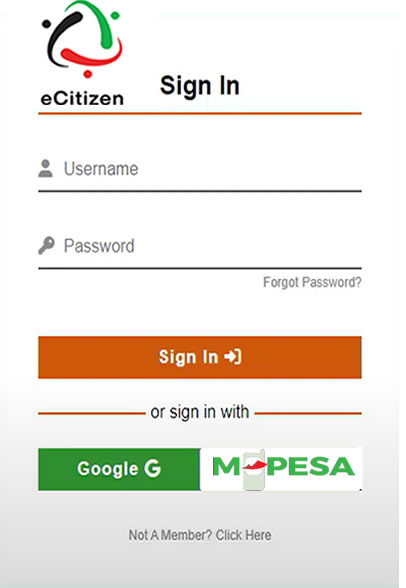}
    \hspace{0.5cm} 
    \includegraphics[scale=0.35]{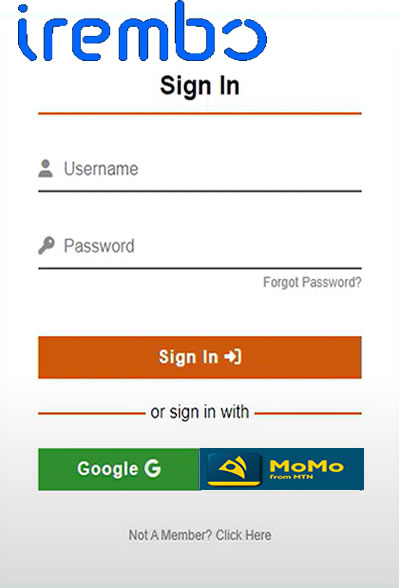}
    \hspace{0.5cm} 
    \includegraphics[scale=0.35]{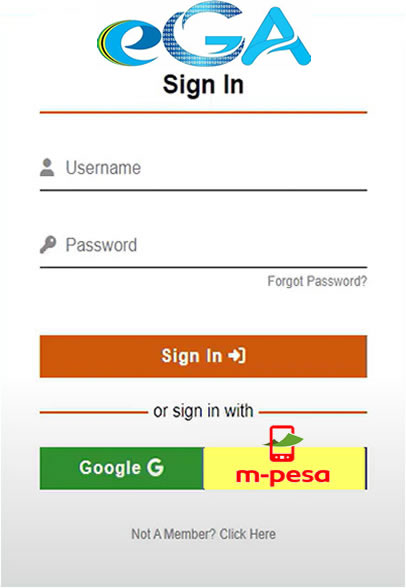}
    \label{fig:interfaces}
   \caption{(a) Login to eCitizen Services Using M-PESA PIN in Kenya. Access services like passport applications, business registration, and land searches.
(b) Access IremboGov Services by Entering MTN MoMo PIN in Rwanda. Apply for IDs, driving licenses, and health insurance conveniently.
(c) Authenticate to e-Government Services Using M-PESA PIN in Tanzania. Pay taxes, register businesses, and manage national ID services online.}
    \label{fig:services}
\end{figure*}

\vspace{-3pt}
The figures in Figure~\ref{fig:services} (a), (b) and (c) are experimental mockups demonstrating potential MMA interfaces for Kenya, Rwanda, and Tanzania, focusing on key usability and security features relevant to East African mobile money applications.
The mockups will generally demonstrate how the interface may look focusing on design principles relevant to East African mobile money applications. They are purported to highlight from a user's standpoint key features which are expected to be in the final project, such as navigation flows, security features, and usability improvements in the localized context for residents of Kenya, Rwanda, and Tanzania.

\vspace{2pt}
\subsection*{\textbf{Comparative Analysis}}
\vspace{-3pt}

We evaluated authentication methods based on accessibility, user experience, security, cost, and scalability. Accessibility is considered the ease of use for users with limited internet connectivity or devices, while user experience focuses on simplicity and intuitiveness for a diverse user base in SSA. Security against threats such as phishing, Man-in-the-Middle attacks, and brute-force attempts was assessed, particularly in resource-constrained settings. Cost evaluation compared user and provider expenses, including infrastructure and operational costs. Scalability was tested to handle increasing user loads and expanding to new services, ensuring suitability for large-scale e-Government platforms. To compare methods, we simulated an e-Government authentication system using mobile money and SSO techniques through a custom-built web application replicating authentication flows. MMA used a Java-based service to simulate USSD interactions, while a standalone authentication server handled user verification, session management, and issued JWT. MongoDB and PostgreSQL store user data, including authentication logs, session details, and simulated Know Your Customer (KYC) information.

\section{Results}
\vspace{-5pt}
In this section, we present the results of our simulations and provide a detailed comparative analysis between  SSO methods and our proposed solution.

\subsection*{\textbf{System Evaluation}}
To quantify the efficacy of our solution, we analyzed three core pillars critical for financial authentication systems:

\vspace{5pt}


\noindent
\subsection{Temporal performance}
\vspace{-3pt}
The delay for each step of the authentication process was measured to ensure that the system met the application requirements in real time. The results are summarized in Table~\ref{tab:delay_metrics}.

\begin{table}[ht!]
\caption{Average Time for Each Step in Authentication.}
\centering
\renewcommand{\arraystretch}{1.6} 
\setlength{\tabcolsep}{5pt} 
\begin{tabular}{|l|c|}
\hline
\textbf{Criteria} & \textbf{Average Time (seconds)} \\
\hline
USSD Push Notification & 1.2 \\
\hline
User PIN Entry & 5.7 \\
\hline
PIN Verification & 0.8 \\
\hline
JWT Generation & 0.3 \\
\hline
\textbf{Total Authentication Time} & \textbf{8.0} \\
\hline
\end{tabular}
\label{tab:delay_metrics}
\end{table}



\subsection{Security robustness}
\noindent
Our assessment combines a review of the cryptographic capabilities implemented with adversarial simulations, addressing two key pillars:

\textit{Validating Encryption:} AES-256 encryption was validated using OpenSSL simulations. The system successfully resisted brute-force attacks and demonstrated compliance with NIST standards \cite{CSRCnist}. Wireshark analysis verified that no plaintext data was transmitted during encrypted communication. A summary of our results can be observed in Table~\ref{tab:security_testing}.

\begin{table}[ht!]
\caption{Security Testing Results for MMA.}
\centering
\renewcommand{\arraystretch}{1.6} 
\setlength{\tabcolsep}{5pt} 
\begin{tabular}{|p{0.4\linewidth}|p{0.5\linewidth}|}
\hline
\textbf{Attack Type} & \textbf{Outcome} \\
\hline
Man-in-the-Middle (MITM) & Prevented using encrypted communication (AES-256). \\
\hline
Brute Force (PIN Guessing) & Account locked after 3 incorrect attempts. \\
\hline
\end{tabular}
\label{tab:security_testing}
\end{table}


\textit{Session Integrity:} We evaluated session continuity under simulated network disruptions (3G/4G signal loss, packet delays) using custom logging tools. The system demonstrated robust recovery mechanisms, maintaining session integrity across 98.5\% of test cases with automated reconnection within 1.8 seconds after network restoration.

\begin{figure}[ht!]
\centering
\includegraphics[width=0.9\linewidth]{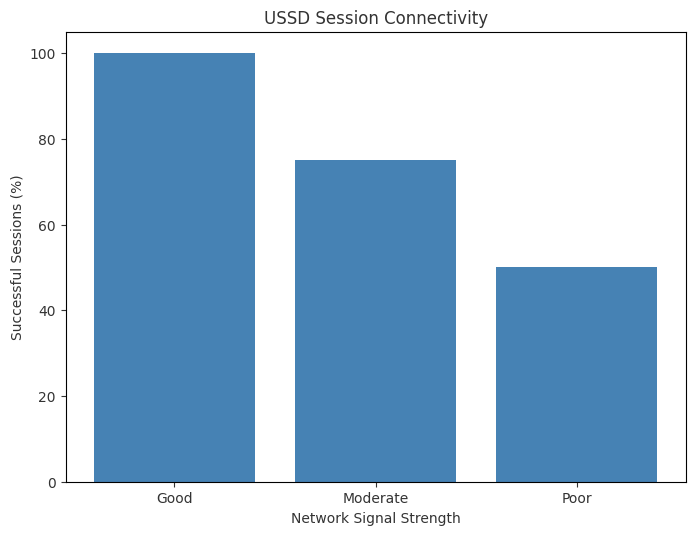}
\caption{Session Integrity Under Varying Network Conditions.}
\label{fig:session_integrity}
\end{figure}

\subsection{Usability}
\noindent
The user interaction was evaluated through pilot testing, focusing on simplicity and efficiency. 
Table~\ref{tab:user_interaction} highlights key metrics. 

\begin{table}[ht!]
\vspace{5pt} 
\caption{User Interaction Metrics for MMA.}
\centering
\renewcommand{\arraystretch}{1.6} 
\setlength{\tabcolsep}{5pt} 
\begin{tabular}{|p{0.6\linewidth}|p{0.3\linewidth}|}
\hline
\textbf{Metric}                     & \textbf{Result} \\
\hline
Average Authentication Time         & 14.5 seconds    \\
\hline
Max Delays under Poor Network       & 20 seconds      \\
\hline
Failed Login Attempts (Lockout)     & Account locked after 3 incorrect PINs. \\
\hline
\end{tabular}
\label{tab:user_interaction}
\end{table}

\noindent
These results confirm that the authentication process is efficient and aligns with the usability design goals.

\begin{table*}[ht!]
    \caption{Comparative Analysis of MMA and Email SSO}
    \centering
    \renewcommand{\arraystretch}{1.6} 
    \setlength{\tabcolsep}{5pt} 
    \begin{tabular}{|p{0.18\linewidth}|p{0.22\linewidth}|p{0.27\linewidth}|p{0.24\linewidth}|}
        \hline
        \textbf{Category} & \textbf{Metric / Aspect} & \textbf{MMA} & \textbf{SSO} \\
        \hline

        \multirow{3}{*}{\textbf{ {Accessibility}}}
        & Internet Requirement & None (USSD-based) & High \\
        \cline{2-4} 
        & Device Compatibility & High (feature phones) & Limited (smart-phones/computers) \\
        \cline{2-4}
        & Low-Literacy Users & High (numeric interface) & Low (text-heavy interface) \\
        \hline

        \multirow{3}{*}{\textbf{ {User Experience (UX)}}}
        & Steps to Authenticate & 2 (USSD, PIN entry) & 3-4 (email, password, 2FA) \\
        \cline{2-4}
        & Familiarity with User Interface & High & Low \\
        \cline{2-4}
        & Language Support & Supports local languages & Limited \\
        \hline

        \multirow{3}{*}{\textbf{ {Security}}}
        & Multi-Factor Auth & Inherent (SIM + PIN) & Optional \\
        \cline{2-4}
        & Phishing Resistance & High & Low \\
        \cline{2-4}
        & Account Recovery & Via MNO (high security) & Email-based (Single Point of Failure) \\
        \hline

        \multirow{3}{*}{\textbf{ {Cost}}}
        & User Cost & Minimal (USSD charges) & Free (requires internet) \\
        \cline{2-4}
        & Implementation Cost & Medium (MNO integration) & Low \\
        \cline{2-4}
        & Maintenance Cost & Low & Low \\
        \hline

        \multirow{3}{*}{\textbf{ {Technical Performance}}}
        & Average Authentication Time & 8.0 seconds & 12-15 seconds \\
        \cline{2-4}
        & Success Rate & 95\% & 80\% (connectivity issues) \\
        \cline{2-4}
        & Security Incident Rate & 0.1\% & 0.5\% \\
        \hline

    \end{tabular}
    \label{tab:mma_sso_comparison}
\end{table*}

\subsection*{\textbf{Comparing MMA with  SSO}}
We present a detailed comparison of MMA and SSO using key criteria: accessibility, user experience, security, cost, and technical performance. Each criterion highlights the strengths and weaknesses of the two approaches, emphasizing MMA's suitability for resource-constrained environments.

\paragraph{Accessibility}
Accessibility evaluates the ease with which users, especially in low-resource settings, can engage with the authentication mechanism. The comparison is summarized in Table~\ref{tab:mma_sso_comparison}.

\noindent
MMA has a clear advantage in accessibility, as it takes advantage of USSD technology to reach users without internet access, supports feature phones, and simplifies interactions for low-literacy users. This makes it particularly suitable for regions with limited digital literacy and connectivity challenges.

\paragraph{User Experience}
User experience evaluates the simplicity, familiarity, and intuitiveness of the authentication process. Table~\ref{tab:mma_sso_comparison} compares key aspects.

\noindent
MMA provides a superior user experience by minimizing the steps required for authentication, relying on familiar interfaces to users, and also supporting local languages.

\paragraph{Security}
\noindent
Security assesses the effectiveness of the system against threats such as phishing, brute force, and session hijacking. Table~\ref{tab:mma_sso_comparison} compares key features.

\noindent
MMA excels in security by providing inherent multi-factor authentication, reducing susceptibility to phishing attacks, and ensuring secure account recovery through MNOs.

\paragraph{Technical Performance}
Technical performance evaluates the efficiency, success rates, and resilience of the authentication process. Table~\ref{tab:mma_sso_comparison} provides a comparison of key metrics.

\noindent
MMA outperforms email-based SSO in three critical metrics:
\begin{itemize}

\item Authentication Speed: 45\% faster (8.0s vs. 12-15s)
\item Success Rate: 15\% higher (95\% vs. 80\%)
\item Security Incident Rate: 80\% lower (0.1\% vs. 0.5\%)
\end{itemize}
These metrics are particularly significant in low-connectivity environments typical of rural SSA, where  authentication methods often fail entirely.

In summary, MMA shows significant promise for e-Government platforms in SSA, particularly with respect to accessibility, security, and user familiarity. However, it faces challenges in cross-platform support and integration complexity compared to  email-based SSO solutions.

\section{Limitations and Future Work}
Despite demonstrating the potential of MMA for e-Government in SSA, this study has limitations. Technically, the reliance on MNO infrastructure creates a single point of failure. USSD, though accessible, lacks strong encryption and is vulnerable to interception and shoulder surfing. Interoperability across platforms is limited, and system performance under extreme load (e.g., during national elections) is untested.

Methodologically, the results are based on simulations rather than real-world deployments. Although diverse profiles were modeled, the lack of field tests with low-literacy and vulnerable users limits generalizability. Our security analysis also did not include red-team exercises or long-term threat modeling.

In terms of implementation, regulatory fragmentation across the SSA complicates adoption. Even low USSD fees can deter users, and integrating with legacy government systems adds complexity.

Future work should prioritize field validation, user studies across diverse demographics, and cost-benefit analyses. Fallback mechanisms (e.g., SMS) should be considered for robustness. Building on works like Binitie et al. \cite{patience2022security, Binitie2024}, research should explore privacy-preserving USSD interfaces to counter shoulder surfing. Finally, usability testing using SUS, heuristic analysis, and cognitive walkthroughs can offer insights into accessibility under real-world constraints.
\section{Conclusion}

This study demonstrates the transformative potential of MMA in improving web platform security, particularly for e-Government services in SSA. Using existing mobile infrastructures and user behaviors, this approach addresses critical gaps in accessibility, user trust, and authentication, offering a viable alternative to  authentication methods in resource-constrained settings. Simulation results showed a 95\% success rate and an average 8-second authentication time under optimal conditions—outperforming  email-based SSO methods, particularly in low-connectivity environments.

Highlights of this work include introducing a threat model tailored to SSA contexts, designing a testable authentication prototype using existing mobile money infrastructures, and presenting comparative usability and performance analysis. The use of SIM and PIN as multifactor authentication mechanisms improved resistance to phishing and brute-force attacks while maintaining usability across diverse device types and network conditions.

Although this work centered on e-Government platforms, the architecture is extensible to sectors like healthcare, education, and financial services. MMA offers a scalable, secure, and inclusive solution that bridges infrastructural gaps in digital service access. Future deployments must consider privacy rights, equitable access, and regulatory oversight to avoid misuse of identity data or exclusion of marginalized users.
This framework lays the groundwork for further innovation in bridging digital divides and improving cybersecurity resilience and acceptance in under-served areas, advancing digital inclusion not only in SSA, but also in other under-served and connectivity limited regions around the world.

\section*{Acknowledgments}
This research was supported in part by the Bill \& Melinda Gates Foundation through the Upanzi Network at Carnegie Mellon University Africa. The authors extend their gratitude to all pilot participants whose feedback informed key aspects of the authentication system's design and evaluation. The views and conclusions expressed herein are solely those of the authors and do not necessarily represent those of the sponsors.

\printbibliography

\end{document}